# Determination of the branching ratio of $CH_3OH + OH$ reaction on water ice surface at 10 K

Atsuki Ishibashi[1][2], Hiroshi Hidaka[2], W. M. C. Sameera[2][3], Yasuhiro Oba[2], Naoki Watanabe[2]

Abstract

The $CH_3O$ and $CH_2OH$ radicals can be important precursors of complex organic molecules (COMs) in interstellar dust. The COMs presumably originating from these radicals were abundantly found in various astronomical objects. Because each radical leads to different types of COMs, determining the abundance ratio of $CH_3O$ to $CH_2OH$ is crucial for a better understanding of the chemical evolution to various COMs. Recent work suggested that the reaction between $CH_3OH$ and OH on ice dust plays an important role in forming $CH_3O$ and $CH_2OH$ radicals. However, quantitative details on the abundance of these radicals have not been presented to date. Herein, we experimentally determined the branching ratio ($CH_3O/CH_2OH$) resulting from the $CH_3OH$ + OH reaction on the water ice surface at 10 K to be 4.3 ± 0.6. Furthermore, the $CH_3O$ product in the reaction would participate in subsequent diffusive reactions even at a temperature as low as 10 K. This fact should provide critical information for COMs formation models in cold molecular clouds.

[1] Corresponding author atsu@lowtem.hokudai.ac.jp
[2] Institute of Low Temperature Science, Hokkaido University N19W8, Kita-ku, Sapporo, Hokkaido 060-0819 Japan
[3] Department of Chemistry, University of Colombo, Colombo, Sri Lanka

1. Introduction

It is commonly thought that COMs are produced by recombination reactions of heavy radical species on interstellar dust grains as the temperature rises with star formation [Garrod & Herbst 2006; Taquet et al. 2012]. In recent years, however, COMs have been detected even in cold molecular clouds where radicals other than H atoms are considered thermally immobile [Öberg et al. 2010; Bacmann et al. 2012; Cernicharo et al. 2012; Vastel et al. 2014; Jiménez-Serra et al. 2016; Soma et al. 2018]. Therefore, several reaction models have been proposed to explain COMs formation in cold regions [Chang & Herbst 2016; Shingledecker et al. 2018; Jin & Garrod 2020]. In either case, the radical species formed on the dust are essential for COMs formation.

The abundance ratio of $CH_3O$ (methoxy) to $CH_2OH$ (hydroxymethyl) radicals on dust is of great interest because it would affect the relative yields of many COMs, such as $HCOOCH_3$ (methyl formate), $CH_3OCH_3$ (dimethyl ether), $CH_3OCH_2OH$ (methoxymethanol), $HCOCH_2OH$ (glycolaldehyde), and $(CH_2OH)_2$ (ethylene glycol), which have been found in astronomical observations. Therefore, quantitative determination of the branching ratio is crucial for constructing better chemical models of COMs formation. Various processes have been proposed for the formation of $CH_3O$ and/or $CH_2OH$ on dust under a low-temperature environment, such as I) hydrogenation of $H_2CO$ [e.g., Watanabe & Kouchi 2002], II) reactions in which the hydrogen atoms of methanol ($CH_3OH$) are abstracted by H atoms, and III) photodecomposition of solid $CH_3OH$ by ultraviolet (UV) photons. Although the branching ratio in case I) has not been determined experimentally, theoretical calculations, including the effect of the water ice surface, have reported that $CH_3O$ formation is dominant over $CH_2OH$ formation [Song & Kästner 2017]. In case II), from deuterium-exposure experiments of $CH_3OH$ on an ice surface, it was suggested that $CH_2OH$ is predominantly formed [Nagaoka et al. 2005, 2007; Hidaka et al. 2009]. The formation of COMs through co-deposition experiments involving H, CO, and CO-hydrogenated species ($H_2CO$, and/or $CH_3OH$), including the reaction processes in cases I) and II), has been extensively demonstrated in previous research [e.g., Fedoseev et al. (2015); Chuang et al. (2016); He et al. (2022)]. These studies reported the formation of many $CH_2OH$-bearing COMs such as $HCOCH_2OH$ and $(CH_2OH)_2$, while $CH_3O$-bearing COMs were less abundant. For case III), the branching ratio was deduced from the types of COMs generated in photolysis experiments of $CH_3OH$ solids [Öberg et al. 2009; Paardekooper et al. 2016;

Yocum et al. 2021; Tenelanda-Osorio et al. 2022]. Chuang et al. (2017) performed experiments co-exposing mixed solids (CO, H$_2$CO, and/or CH$_3$OH) to UV and H atoms under various UV/H atom flux ratios, in which the above three radical formation processes can occur. However, their experiments showed that the relative yields of CH$_3$O-bearing COMs (i.e., HCOOCH$_3$) to CH$_2$OH-bearing COMs (i.e., HCOCH$_2$OH and (CH$_2$OH)$_2$) are significantly low in contradiction to astronomical observations [Chuang et al. 2017]. This suggests that reaction processes other than the abovementioned ones are decisively lacking.

Recently, the direct observation of an ice dust analog using transmission electron microscopy indicated that CO solid and the products of successive CO hydrogenation reactions would not fully cover the water ice mantles of dust even in the CO freezing phase [Kouchi et al. 2021a, b]. Under such a surface structure, CH$_3$OH can exist next to water molecules; thus, chemical processes with H$_2$O would play an important role. Our previous photolysis experiments of CH$_3$OH adsorbed on amorphous solid water (ASW) surface showed the abundant formation of CH$_3$OCH$_2$OH and HCOOCH$_3$ derived from CH$_3$O compared to HCOCH$_2$OH and (CH$_2$OH)$_2$ derived from CH$_2$OH [Ishibashi et al. 2021], which is qualitatively consistent with their astronomical observations [Jørgensen et al. 2020; Mininni et al. 2020; Manigand et al. 2020; El-Abd at al. 2019; Tercero et al. 2018; Rivilla et al. 2017; Coutens et al. 2015; Taquet et al. 2015]. Ishibashi et al. suggested that the reaction of CH$_3$OH and OH formed from photodecomposition of water ice effectively provides CH$_3$O radical, leading to the predominant formation of CH$_3$O-bearing molecules. That is, reaction (1) is more prevailing than reaction (2).

$$\text{CH}_3\text{OH} + \text{OH} \xrightarrow[\text{On ASW}]{k_1} \text{CH}_3\text{O} + \text{H}_2\text{O} \quad (1)$$

$$\xrightarrow[\text{On ASW}]{k_2} \text{CH}_2\text{OH} + \text{H}_2\text{O} \quad (2)$$

Ishibashi et al. (2021) demonstrated that the branching ratio of reactions (1) and (2) significantly affects the abundance ratio of CH$_3$O to CH$_2$OH radicals on dust. However, there is currently no quantitative information on the branching ratio in reactions (1) and (2).

In this study, we determined the branching ratio (CH$_3$O/CH$_2$OH) in the reaction of CH$_3$OH and OH radicals on the ASW surface at 10 K by employing the Cs$^+$ ion pickup method, a high-sensitivity surface analysis method [Kang 2011; Ishibashi et al. 2021]. This innovative method provides 2–3 orders of magnitude higher sensitivity for detecting adsorbates than conventional methods such as Fourier transform infrared spectroscopy, enabling in situ observation of even radical species with low abundance on the surface.

## 2. Experimental

Experiments were conducted using an apparatus designed for the surface reaction experiments, which was equipped with a highly sensitive detection system based on the $Cs^+$ ion pickup method as described elsewhere [Ishibashi et al. 2021]. The ASW (~10 ML) was prepared by background vapor deposition of $H_2O$ onto an aluminum substrate at 30 K, followed by generating OH radicals through exposing ASW for 2–15 min to UV photons (~$6 \times 10^{12}$ photons $cm^{-2}$ $s^{-1}$) from a conventional deuterium lamp (115-400 nm). The photons from the lamp photodissociate $H_2O$ mainly into H + OH with minor channels, $H_2$ + O and 2H + O [Slanger & Black 1982]. However, the most of these volatile products, especially H atom, on the surface immediately desorb upon photodissociation at 30 K. In the present experiment, to avoid possible effects of such volatiles, the deposition and exposure temperature was set to 30 K. The coverage of OH on the ASW was estimated to be an order of magnitude of 0.01 from a signal intensity of OH radical (see Section 3.1). After cooling ASW to 10 K, the $CH_3OH$ gas (Kanto chemical, purity 99.8 %) was then deposited through a microcapillary plate with a different gas line for $H_2O$ at a rate of ~$1.3 \times 10^{11}$ molecules $cm^{-2}$ $s^{-1}$ onto the UV-irradiated ASW. The total amount of deposited methanol was estimated to be approximately $2.0 \times 10^{14}$ molecules $cm^{-2}$ over 20 minutes using a reflection-type FTIR measurement. The present experiments also used methanol isotopologues, i.e., $CD_3OH$ (Acros organics, 99.5 atom % D) and $CH_3OD$ (Sigma-Aldrich, 99.5 atom % D). According to the coverage of OH and the amount of $CH_3OH$, the kinetic energy of gaseous $CH_3OH$ was expected to be sufficiently lost to the thermal energy level before colliding with OH. The $Cs^+$ ion injection energy for monitoring the surface composition is about ~17 eV, which enable us to nondestructively monitor the reactant and product species (Mass X) on the surface. The mass numbers of the picked-up species are identified by mass analysis with a quadrupole mass spectrometer without an ionization cell as a mass number of 133 + X. In the present experiments, the $Cs^+$ ion bombardment did not affect the reaction system (e.g., radical destruction, diffusion induction), as demonstrated in Appendix A1.

To clarify the details of reaction paths, we also used deuterated isotopologues of methanol. It is important to prevent undesired alterations in the degree of deuteration of the sample gases which can be caused by the deuterium–hydrogen (D-H) substitution reaction with $H_2O$ adsorbed on the metal inner wall of the gas line systems. In our setup, the gas line with a microcapillary plate used for methanol deposition was isolated from the main chamber by a gate valve which was closed during ASW preparation. This arrangement minimized the $H_2O$ contamination in the gas line for isotopologues. A few weeks before changing the sample isotopologue for the deposition, the gas line system was baked out at 60℃ for over one week, and the isotopologues were circulated via a pre-gas flow. These approaches enabled the isotopologues, even $CH_3OD$, to be deposited with a purity of >90%.

To ensure a constant deposition rate, the opening of the variable leak valve was fixed during all experiments, while the gas flow was controlled by opening and closing the gate valve. The back pressures of all the isotopologues were set to the same value, which was measured by an absolute pressure transducer mounted at the gas cell.

## 3. Results and Discussions

### 3.1. Reaction Pathways of Products Formed by Methanol Deposition to UV-exposed Amorphous Solid Water (ASW)

#### 3.1.1. Detected Products after Methanol Deposition

Fig. 1(a) shows the mass spectra of ASW vapor-deposited at 10 K before and after UV irradiation for 15 min. The formation of OH radicals was confirmed after UV irradiation by the increase in the peak intensity at Mass 151, which was the sum of Masses 133 ($Cs^+$) and 17 (OH). Assuming the same pickup efficiency for $H_2O$ and OH, the surface abundance of OH is estimated to be ~1% relative to $H_2O$. This abundance agrees with the result of previous study that estimated based on the UV irradiation time and photodissociation cross section of $H_2O$ [Miyazaki et al. 2022]. Similarly, the formation of $O_2$ (Mass 32(165)), $HO_2$ (Mass 33(166)), and $H_2O_2$ (Mass

34(167)) were confirmed. Fig. 1(b) shows the pickup spectra obtained in the experiments of $CH_3OH$ (Mass 32(165)) deposited on nonirradiated and UV-irradiated ASW, respectively. Carbon-bearing species, such as $CH_3O$ and/or $CH_2OH$ (Mass 31(164)), and $H_2CO$ (Mass 30(163)) were observed only for the UV-irradiated ASW, indicating that they were produced by the interactions of the methanol with the photoproducts of ASW such as OH. Since $CH_3O$ and $CH_2OH$ cannot be separated by mass analysis, we also performed isotope-labeling experiments using methanol isotopologues $CD_3OH$ (Mass 35(168)) and $CH_3OD$ (Mass 33(166)) (hereafter, these experiments are denoted as "$X$ experiment," where $X$ represents $CH_3OH$, $CD_3OH$, or $CH_3OD$). Figs. 1(c) and (d) show the pickup spectra obtained in $CD_3OH$ and $CH_3OD$ experiments on nonirradiated and UV-irradiated ASW, respectively. On the nonirradiated ASW, different isotopologues from deposited methanol (i.e., $CHD_2OH$ in Fig. 1(c) and $CH_3OH$ in Fig. 1(d)) and HDO were detected, which may have originated from the impurity in the chemical reagent and/or might have been formed by the deuterium–hydrogen (D–H) exchange with water at the gas line before the methanol deposition. Thus, those isotopologues contaminants will also exist on UV-irradiated ASW. On the UV-irradiated ASW, mass peaks corresponding to the reaction products of deuterated counterparts of $CH_3O$, $CH_2OH$, and $H_2CO$ were observed.

Those peak patterns indicate the formation of methoxy and hydroxymethyl radicals occur. However, unfortunately, those reaction products sometimes overlap with the contaminants and some photoproducts of ASW (i.e., $O_2$, $HO_2$, and $H_2O_2$). Therefore, we carefully corrected the signal intensities of reaction products to eliminate the contributions of contaminants and photo products according to the methods shown in Appendix A2. The reaction pathways in the experiments using each methanol isotopologue were identified by monitoring the time variations of reactants and products, as shown in the following section.

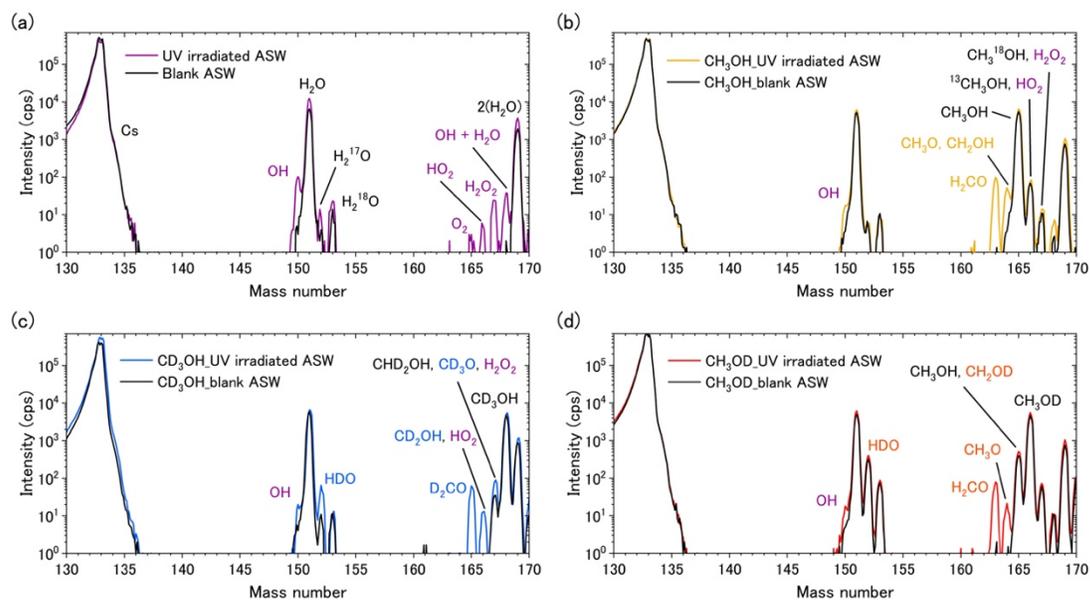

Figure 1. (a) Pickup spectra of ASW before (black) and after (purple) exposure to UV for 15 min. (b-d) Pickup spectra of (b) $CH_3OH$ (yellow), (c) $CD_3OH$ (blue), and (d) $CH_3OD$ (red) deposited on UV-irradiated ASW. Pickup spectra of these methanol isotopologues deposited on nonirradiated ASW are also shown in each panel for comparison (black).

3.1.2. The Time Evolution of Products during Methanol Deposition

Fig. 2(a,b) shows the time evolution of the signal intensities for the reactants and products picked up during the deposition of $CH_3OH$ onto 15 min UV pre-irradiated ASW. The $CH_3OH$ signal increased with the $CH_3OH$ deposition time, whereas the preproduced OH signal decreased. Furthermore, the Mass 31 signal corresponding to $CH_3O$ and/or $CH_2OH$ increased, and then the Mass 30 signal corresponding to $H_2CO$ increased sequentially. This result clearly shows that, even at 10 K, the reaction of $CH_3OH$ + OH occurred on ASW, followed by $H_2CO$ formation. Note that the Y-axis in Fig. 2(b) represents the observed intensities corrected for quantitative analysis (see Appendix A3 for correction details). Hereafter, all experimental data displayed in the figures were corrected in the same manner. Fig. 2(c-e) shows the results of experiments using $CH_3OH$ isotopologues ($CD_3OH$ and $CH_3OD$) for obtaining isotope-labeling radicals. The methoxy and hydroxymethyl radicals were detected as $CD_3O$ (Mass 34) and $CD_2OH$ (Mass 33); $CH_3O$ (Mass 31) and $CH_2OD$ (Mass 32) for the $CD_3OH$ (Fig. 2(c)) and $CH_3OD$ (Fig. 2(d)) experiments, respectively. In addition, signal intensities of formaldehyde products at Mass 32 ($D_2CO$) in the $CD_3OH$ experiment and Mass 30 ($H_2CO$) in the

CH$_3$OD experiment also increased with a similar trend as H$_2$CO in the CH$_3$OH experiment (Fig. 2(e)). Note that although the signal intensity of each adsorbate is proportional to its number density on the surface, the proportional constants are not identical due to difference in pickup efficiency.

The experiments using methanol isotopologues revealed the occurrence of the following reactions:

・CD$_3$OH experiments

$$CD_3OH + OH \xrightarrow[\text{On ASW}]{k_3} CD_3O + H_2O \quad (3)$$

$$CD_3OH + OH \xrightarrow[\text{On ASW}]{k_4} CD_2OH + HDO \quad (4)$$

・CH$_3$OD experiments

$$CH_3OD + OH \xrightarrow[\text{On ASW}]{k_5} CH_3O + HDO \quad (5)$$

$$CH_3OD + OH \xrightarrow[\text{On ASW}]{k_6} CH_2OD + H_2O \quad (6)$$

These results indicate that the Mass 31 signal in the CH$_3$OH experiment (Fig. 2(b)) included both CH$_3$O and CH$_2$OH contributions.

Formaldehyde is considered to be formed by the following reactions:

・Sequential reaction of methoxy radical

$$CH_3O + OH \xrightarrow[\text{On ASW}]{k_7} H_2CO + H_2O \quad (7)$$

$$CD_3O + OH \xrightarrow[\text{On ASW}]{k_8} D_2CO + HDO \quad (8)$$

・Sequential reaction of hydroxymethyl radical

$$CH_2OH + OH \xrightarrow[\text{On ASW}]{k_9} H_2CO + H_2O \quad (9)$$

$$CD_2OH + OH \xrightarrow[\text{On ASW}]{k_{10}} D_2CO + H_2O \quad (10)$$

$$CH_2OD + OH \xrightarrow[\text{On ASW}]{k_{11}} H_2CO + HDO \quad (11)$$

Thus, the formation of formaldehyde from methanol requires two OH radicals. Because the surface coverage of OH is estimated to be ~0.01, the occurrence of reactions (7)-(11) on the ice surface suggests that methoxy and/or hydroxymethyl encounter OH radicals at 10 K via diffusion. It should be noted that Cs⁺ ion irradiation does not contribute to the chemical reactions described above, as shown in Appendix A1.

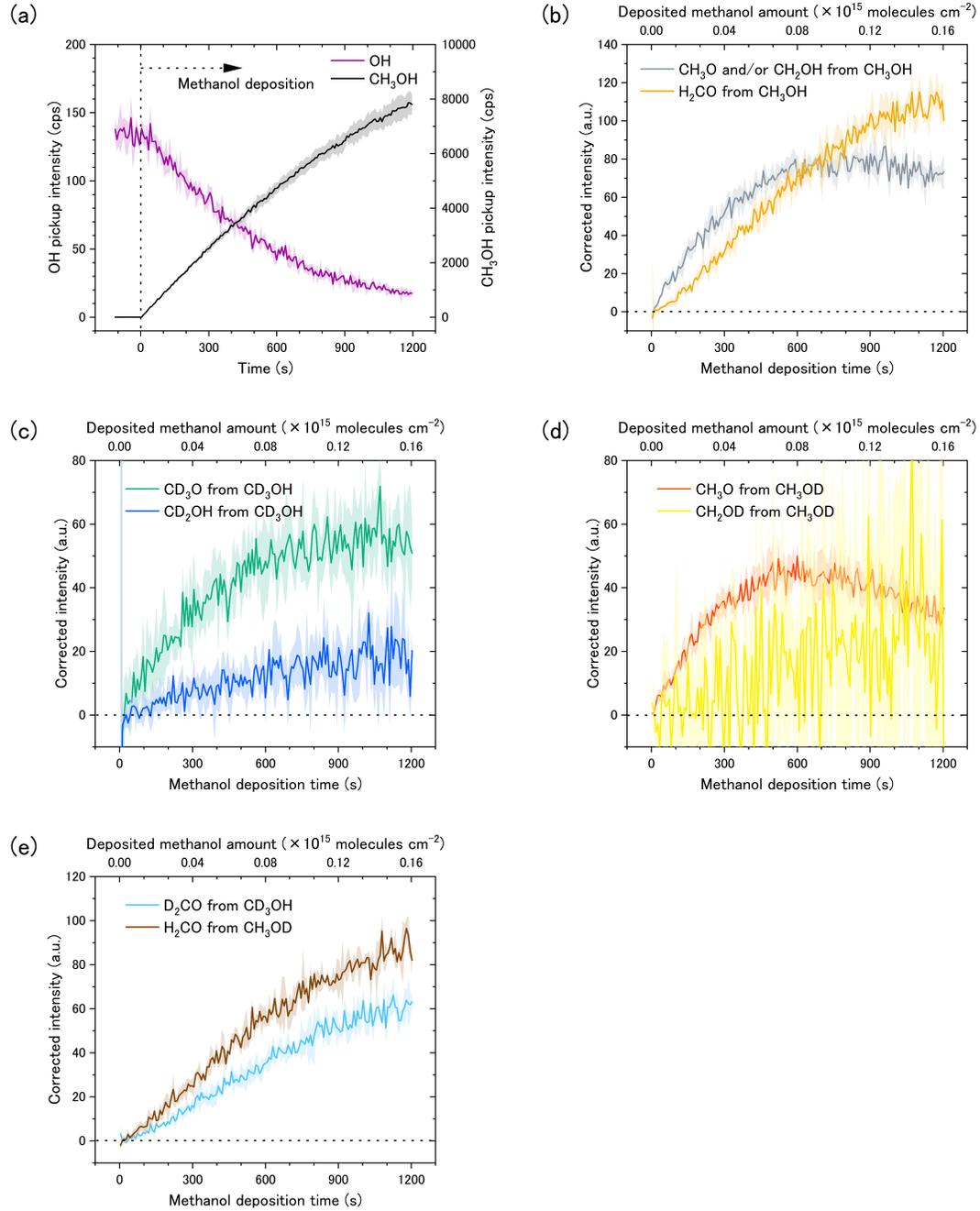

Figure 2. Time evolution of the signals of surface species at 10 K during CH$_3$OH or its isotopologues (CD$_3$OH or CH$_3$OD) deposition on the ASW surface bearing OH radicals preproduced via UV irradiation for 15 min, (a) Mass 17: OH (purple) and Mass 32: CH$_3$OH (black). The region after the dotted vertical lines (0 s) indicates the duration of CH$_3$OH deposition at a rate of ~1.3 × 10$^{11}$ molecules cm$^{-2}$ s$^{-1}$, as estimated using Fourier transform infrared spectroscopy measurements. (b) Signal intensities of Mass 30: H$_2$CO (orange) and Mass 31: CH$_3$O and/or CH$_2$OH (ash). (c) Mass 34: CD$_3$O (green) and Mass 33: CD$_2$OH (blue) produced from CD$_3$OH deposition experiments. (d) Mass 31: CH$_3$O (red) and Mass 32: CH$_2$OD (yellow) produced from CH$_3$OD deposition experiments. (e) Mass 32: D$_2$CO (light blue) and Mass 30: H$_2$CO (brown), obtained from CD$_3$OH and CH$_3$OD experiments, respectively. The solid lines are the means of three different measurements, and the shadows represent statistical errors.

### 3.1.3. Confirmation of Reaction Path

To confirm the above reaction pathways, we measured the [OH]$_0$ (initial number density of OH radicals before methanol deposition) dependence of the pickup signal intensities of products by performing methanol vapor deposition experiments with different [OH]$_0$, which was controlled by varying the UV irradiation time before methanol deposition. Fig. 3 shows the correlation between signal intensity for [OH]$_0$ and UV irradiation time for 2, 3, 5, 7, 10, and 15 min. From reactions (1)-(6), the methoxy and hydroxymethyl radicals are produced from reactions with one OH. In contrast, from reactions (7)-(11), two OH radicals are required for formaldehyde (D$_2$CO and H$_2$CO) production. Therefore, the yields of products from former and latter reactions should have a different dependence of [OH]$_0$. The evolution of number densities for the reaction products with the CH$_3$OH deposition time assumed from the above reaction pathways (CH$_3$OH experiment as an example) was approximated as follows (the derivation of the equation is found in Appendix A4).

$$[\text{CH}_3\text{O}]_t = \frac{k_1}{k_{1+2}}(1 - e^{-k_{1+2}t})[\text{OH}]_0 - \frac{k_1 k_7}{2k_{1+2}^2}(1 - e^{-k_{1+2}t})^2[\text{OH}]_0^2 \quad (12)$$

$$[\text{CH}_2\text{OH}]_t = \frac{k_2}{k_{1+2}}(1 - e^{-k_{1+2}t})[\text{OH}]_0 - \frac{k_2 k_9}{2k_{1+2}^2}(1 - e^{-k_{1+2}t})^2[\text{OH}]_0^2 \quad (13)$$

$$[\text{H}_2\text{CO}]_t = \frac{k_1 k_7 + k_2 k_9}{2k_{1+2}^2}(1 - e^{-k_{1+2}t})^2[\text{OH}]_0^2 \quad (14)$$

where $k_{1+2} = k_1 + k_2$. Similar equations can be written for the isotopologues by choosing an appropriate $k$. When formaldehyde is produced in Eq. (14), the amount produced at a specific time should quadratically correlate with [OH]$_0$.

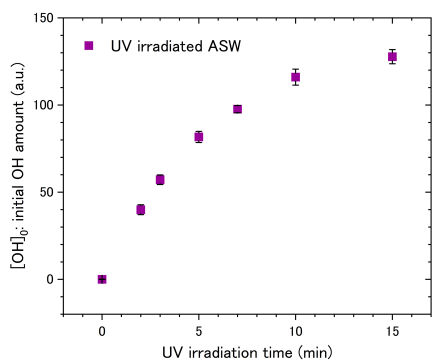

Figure 3. Temporal variation of OH pickup intensities, $[OH]_0$, upon UV irradiation before methanol deposition. Each error bar is statistical derived from measurements more than 6 times.

Fig. 4 shows the pickup signal intensities of methoxy, hydroxymethyl, and formaldehyde at 240 s for the (a) $CD_3OH$ and (b) $CH_3OD$ experiments. The variations of $D_2CO$ and $H_2CO$ intensities were found to be quadratic of $[OH]_0$. Thus, the correlations shown in Fig. 4 are strong evidence that formaldehyde is produced in reactions represented by Eq. (14). According to Eqs. (12) and (13), the variation in the number densities of methoxy and hydroxymethyl yields include the square terms of $[OH]_0$, which represent consumption by the sequential reactions. Nevertheless, the $CD_3O$, $CD_2OH$, and $CH_3O$ intensities increased approximately linearly with $[OH]_0$. The $CH_2OD$ intensity also appears to linearly increase, albeit with large errors due to the significant influence of the contaminant ($CH_3OH$) (see Appendix A2). That is, the contributions of consumption (i.e., formation of formaldehyde) of reactions (8, 7) and (10, 11) are much smaller than those of generation from reactions (3, 5) and (4, 6), respectively. It should be noted that although the intensities of formaldehyde were high, it does not indicate its significant yield. The high detected intensities of formaldehyde can be attributed to the higher pickup efficiency of formaldehyde than those of reactants, namely methoxy and/or hydroxyl radicals.

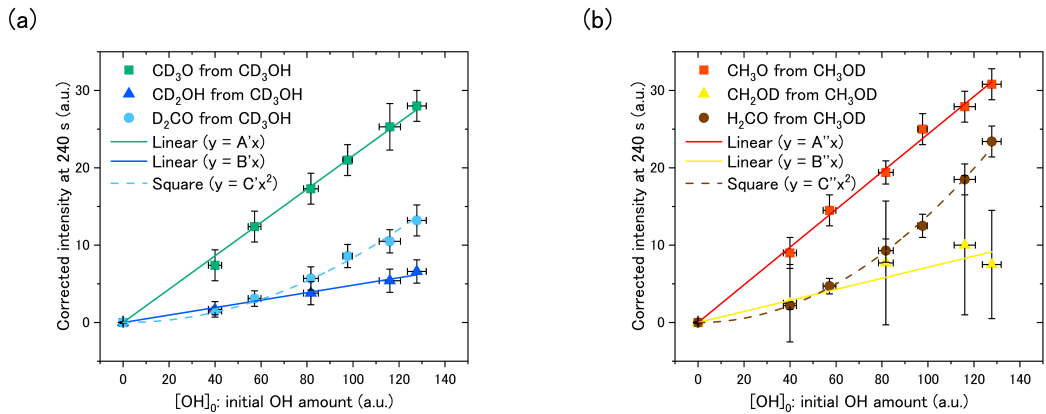

Figure 4. $[OH]_0$ dependence of pickup signal intensities at 240 s after $CH_3OH$ vapor deposition begins. (a) $CD_3O$, $CD_2OH$, and $D_2CO$ generated in the $CD_3OH$ experiments. $CD_3O$ and $CD_2OH$ have a linear dependence, and $D_2CO$ has a square dependence on $[OH]_0$. (b) $CH_3O$, $CH_2OD$ and $H_2CO$ produced in the $CH_3OD$ experiments. $CH_3O$ has a linear dependence, and $H_2CO$ has a square dependence on $[OH]_0$. $CH_2OD$ appears to represent a linearly dependence, although the error is large. Each plot and error bar was obtained from the results of three experiments. The UV irradiation times were 2, 3, 5, 7, 10, and 15 min.

### 3.2. Determination of the Branching Ratio in reaction between Methanol and OH Radical on ASW

We used methanol isotopologues ($CD_3OH$ and $CH_3OD$) in the present experiment to study the reactions of methanol with OH radicals on ASW at the functional group level. However, if isotopic effects appear in the rate constants of the chemical reactions, each isotopologue experiment may lead to different branching ratios. Therefore, to accurately estimate the branching ratio of reactions (1) and (2) in experiments using isotopologues, it was necessary to clarify if isotopic effects existed in the reactions.

Fig. 5(a) shows the variations in the signal intensities of $CD_3O$ and $CH_3O$ in the $CD_3OH$ and $CH_3OD$ experiments, respectively, with deposition time in the region where the contributions of sequential reactions should be small (0–400 s). In each isotopologue experiment, there was little difference between the peak intensities of $CD_3O$ and $CH_3O$ (H or D abstraction from OH or OD, respectively). In addition, there was no detectable difference in the pickup signal intensities between $CD_2OH$ and $CH_2OD$ (D or H abstraction from $CD_3$ or $CH_3$, respectively) (Fig. 5(b)). If the pickup efficiencies of the radicals were approximately the same between the isotopologues, we can conclude that the branching ratio obtained in each isotopologue experiment should not be affected by the isotopic effect on the H (D) abstraction reactions. This assumption seems reasonable because the pickup efficiencies between methanol isotopologues did not vary substantially (See Section 3.3 for a discussion of this cause). The discrepancy in the pickup intensities above ~500 s between $CD_3O$ in Fig. 2(c) and $CH_3O$ in Fig. 2(d) is likely due to isotopic effects on the consumption rate of methoxy radicals through the sequential reactions (8) and (7) (see Section 3.4 for detail). Therefore, the branching ratio ($CH_3O/CH_2OH$) was estimated under the reasonable assumption that the same branching ratios would be obtained in each isotopologue experiment.

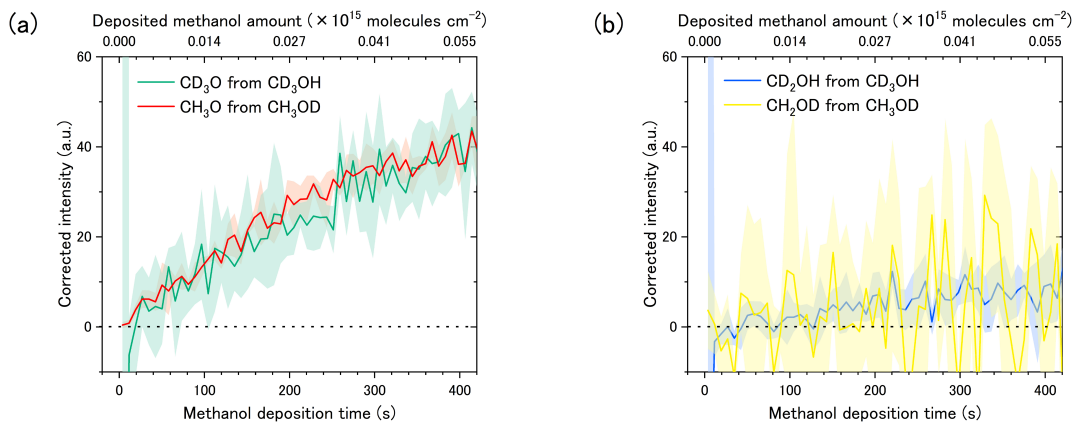

Figure 5. (a) Variations in the pickup signal of $CD_3O$ (green) and $CH_3O$ (red) generated in the $CD_3OH$ and $CH_3OD$ experiments, respectively. (b) Variations in the pickup signal of $CD_2OH$ (blue) and $CH_2OD$ (yellow) generated in the $CD_3OH$ and $CH_3OD$ experiments, respectively. The signal is the mean value of three experiments and the shaded area represents the error. The horizontal dotted line represents zero of the corrected pickup intensity.

In a simple analysis, a branching ratio of ~4 was obtained using the pickup signal intensity ratio of $CD_3O$ and $CD_2OH$ produced in reactions (3) and (4), respectively, or that of $CH_3O$ and $CH_2OD$ produced in reactions (5) and (6), respectively, from Fig. 5. However, this idea cannot be adopted because (i) methoxy and hydroxymethyl radicals may have different pickup efficiencies, and (ii) these radicals are further consumed in reactions (7), (8), (10), and (11). Instead, to derive the branching ratio of these radicals, we used the HDO signals generated as by-products of $CH_3O$ from reaction (5) in the CH$_3$OD experiments and CD$_2$OH from reaction (4) in the CH$_3$OH experiments. As shown in Fig. 6(a), HDO should be obtained in the same amount as CH$_3$O or CD$_2$OH produced in the experiments with each methanol isotopologue. Additionally, HDO is not consumed in sequential reactions. Therefore, the use of the HDO pickup signal as a probe is appropriate for estimating the reaction branching ratio. HDO is also produced in sequential reactions (8) and (11) in which formaldehyde is produced. However, the contribution of the HDO yields from reactions (8) and (11) can be separated from those from reactions (4) and (5) because the HDO yields from former and latter reactions should have the different dependence of [OH]$_0$, as shown in Fig. 4. From Eq. (12)-(14), the number density of HDO at a specific time "$t_a$" in each isotopologue experiment were described as following equations, respectively, under the above reasonable assumption of $k_1 = k_3 = k_5$ and $k_2 = k_4 = k_6$.

· CH$_3$OD experiments

$$[\text{HDO}_{\text{CH}_3\text{OD}}]_{t=t_a} = \frac{k_1}{k_{1+2}}(1 - e^{-k_{1+2}t_a})[\text{OH}]_0 + \frac{k_2 k_{11}}{2k_{1+2}^2}(1 - e^{-k_{1+2}t_a})^2[\text{OH}]_0^2$$

$$= A \times [\text{OH}]_0 + D \times [\text{OH}]_0^2 \quad (15)$$

· CD$_3$OH experiments

$$[\text{HDO}_{\text{CD}_3\text{OH}}]_{t=t_a} = \frac{k_2}{k_{1+2}}(1 - e^{-k_{1+2}t_a})[\text{OH}]_0 + \frac{k_1 k_8}{2k_{1+2}^2}(1 - e^{-k_{1+2}t_a})^2[\text{OH}]_0^2$$

$$= B \times [\text{OH}]_0 + C \times [\text{OH}]_0^2 \quad (16)$$

In these equations, the first term refers to the former reaction, and the second term refers to the latter reaction. Note that the second terms in Eqs. (12) and (13) are not necessary to be considered in HDO productions, because HDO is not consumed by sequential reactions unlike radicals, as mentioned above. Therefore, by fitting the pickup signal intensity of HDO at a specific time relative to [OH]$_0$ with a quadratic function, the contribution of HDO from the latter reactions (8) and (11) can be evaluated.

We measured the [OH]$_0$ dependence of the pickup signal intensities of HDO similar to Fig. 4. Fig. 6(b) shows the [OH]$_0$ dependence of the HDO pickup intensity at the methanol deposition time of 240 s. The contaminations were removed from the plotted HDO intensity (see Appendix A2 for details). The HDO from reactions (4) and (8) obtained in the CD$_3$OH experiments (blue inverted triangles) tended to increase quadratically with respect to [OH]$_0$, whereas the HDO from reactions (5) and (11) in the CH$_3$OD experiments (red squares) increased almost linearly with $[OH]_0$. This is likely due to the negligible contribution of reaction (11) (i.e., the second term in Eq. (15)). Therefore, the $[OH]_0$ dependence of HDO obtained from the $CH_3OD$ experiments was fitted with a linear function (y = Ax) by neglecting the contribution of reaction (11) (See Section 3.4 for validity), and that obtained from the $CD_3OH$ experiments was fitted with a quadratic function (y = Bx + Cx$^2$). From the obtained linear component ratio (A/B), the $CH_3O/CH_2OH$ branching ratio was calculated as 4.3 ± 0.6. Even when the data points are fitted with the quadratic function, the derived branching ratio of 3.8 ± 0.8 is consistent to the above value within the error. In addition, the values for the chi-square degrees freedom in these fittings are almost equivalent.

In addition, the first term of Eq. (15) (i.e., A × $[OH]_0$) refers to the total amount of $CH_3O$ produced by reaction (5), which is equal to the total amount of $CD_3O$ produced by reaction (3), and the second term of Eq. (16) (i.e., C × $[OH]_0^2$) refers to the amount of $CD_3O$ consumed by reaction (8). Therefore, the relation of consumption to production "C/A × $[OH]_0$" indicated that ~14% of the produced $CD_3O$ radicals were consumed by the sequential reaction (8) at 240 s under the condition of ~1% OH abundance (i.e., $[OH]_0$ ~ 130 in the 15 min UV-exposed ASW experiment). This is also consistent with the fact that in Fig. 4, the contribution of consumption by sequential reactions of methoxy radicals is small and the yields of methoxy radicals can be approximately linear with respect to $[OH]_0$ within the error range.

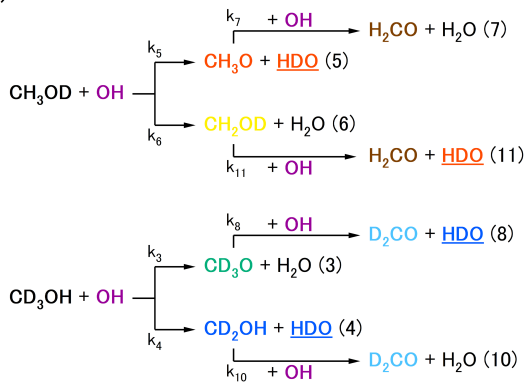

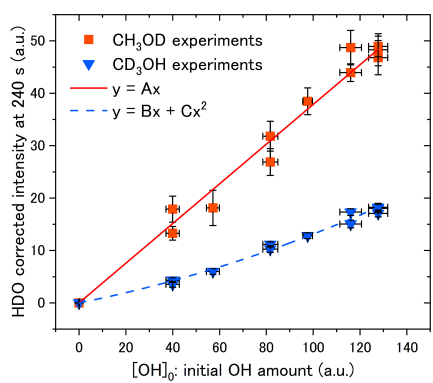

Figure 6. (a) Reaction scheme for the $CH_3OD$ and $CD_3OH$ experiments. (b) $[OH]_0$ dependence of HDO intensity at 240 s generated in the $CH_3OD$ (red square) and $CD_3OH$ (blue inverted triangle) experiments. We fitted HDO from the $CH_3OD$ experiments with $y = Ax$ (solid line) because of the negligible contribution of reaction (11), and HDO from the $CD_3OH$ experiments with $y = Bx + Cx^2$ (dashed line). The obtained fitting values are $A = 0.379 \pm 0.008$, $B = 0.089 \pm 0.010$, and $C = 0.00042 \pm 0.00009$. From the ratio of the linear components (A/B) obtained from the fitting, we estimated the $CH_3O/CH_2OH$ branching ratio to be $4.3 \pm 0.6$. The UV irradiation times are 2, 3, 5, 7, 10, and 15 min. Each plot and error bar was obtained from the results of three UV experiments, with two blank experiments performed before and after the UV experiments. In each of the $CH_3OD$ and $CD_3OH$ experiments, there are three plots at UV 15 min, two plots at UV 10, 5, and 2 min, and one plot at UV 7 and 3 min. The fact that multiple plots at the same $[OH]_0$ show similar values indicates that contamination is correctly removed by the blank experiments no matter when the experiment is performed.

### 3.3 Comparison with Gas Phase Reaction

In the gas phase, the $CH_3OH + OH$ reaction has been extensively studied both experimentally and theoretically [Xu & Lin 2007; Shannon et al. 2013; Gomez Martín et al. 2014; Antiñolo et al. 2016; Gao et al. 2018; Roncero et al. 2018; Nguyen et al. 2019; Ocaña et al. 2019; del Mazo-Sevillano et al. 2019]. The branching ratio of this reaction is pressure dependent, and the decisive value for the branching ratio in the low-pressure limit environment, which is important in gas-phase interstellar chemistry, is currently under discussion [Canosa 2019]. Meanwhile, Gao et al. calculated that the branching ratio of the $CH_3OH + OH$ reaction under the high-pressure limit (HPL) conditions was almost 100% $CH_3O$ at temperatures <80 K. In the HPL calculations, since the formed pre-reactive complex (PRC) rapidly relaxes energy (fully equilibrated and thermalized) with third body collisions before the abstraction reaction, the reaction proceeds by tunneling in the ground state of the PRC at very low temperatures. Therefore, the branching ratio is determined by the difference between the rate constants of tunneling reactions toward $CH_3O$ and $CH_2OH$ formations, which depend on the shape of the potential energy barriers of each reaction [Gao et al. 2018]. Intuitively, reactions on the ASW surface might be similarly to those in HPL conditions because the PRC formed on ASW can rapidly dissipate energy at the surface.

However, the reaction branching ratio on the ASW surface obtained in this study was not extremely biased toward $CH_3O$, unlike the HPL gas-phase results. The difference in the branching ratio is likely due to the barrier shape of the CH$_3$OH + OH reaction on the ASW surface. According to previous quantum chemical calculations, the involvement of one or two H$_2$O molecules changes the barrier shape of the reaction between CH$_3$OH and OH and forms multiple PRC structural isomers with corresponding multiple transition states [Jara-Toro et al. 2017; Chao et al. 2019; Wu et al. 2020]. At the cold surface, multiple PRC structural isomers involving H$_2$O would rapidly dissipate energy to the surface after their formation; therefore, they would be unable to overcome their isomerization barrier to obtain the most stable structure and should be able to exist stably with different structures. In addition, on the ASW surface, more H$_2$O molecules will be involved in the PRC composition; hence, it can be inferred that more multiple PRC structural isomers will form depending on the adsorption sites and orientations of CH$_3$OH and OH as shown in left side of Fig. 7. In fact, CH$_3$OH and OH radicals are known to exist in various adsorption states on the ASW surface [Miyazaki et al. 2020; Ferrero et al. 2020]. In addition, most recently, various reaction barriers depending on the adsorption structure of CH$_3$OH and OH on ASW were calculated for their reaction [Sameera et al. 2023]. In that case, the rate constant ratio (i.e., $k_1/k_2$) has a unique value depending on each PRC structure on ASW surface. Therefore, the branching ratio obtained in the present experiments should be average of the branching ratios obtained from each of the multiple PRC structural isomers on ASW (just denoted by PRC $\alpha$, PRC $\beta$, and PRC $\gamma$ in right side of Fig. 7). The predominant formation of CH$_3$O is qualitatively understandable, considering that bond formation between CH$_3$OH and H$_2$O molecules occurs preferentially on the hydroxyl group side of CH$_3$OH, such as PRC β and γ, under H$_2$O abundant conditions at low temperatures [Dawes et al. 2016]. That is, the OH radical adsorbed on ASW tends to interact with the hydroxyl side of CH$_3$OH. In this context, the adsorption orientations and structures of reactants may be important factors influencing the reaction branching on the ice surface.

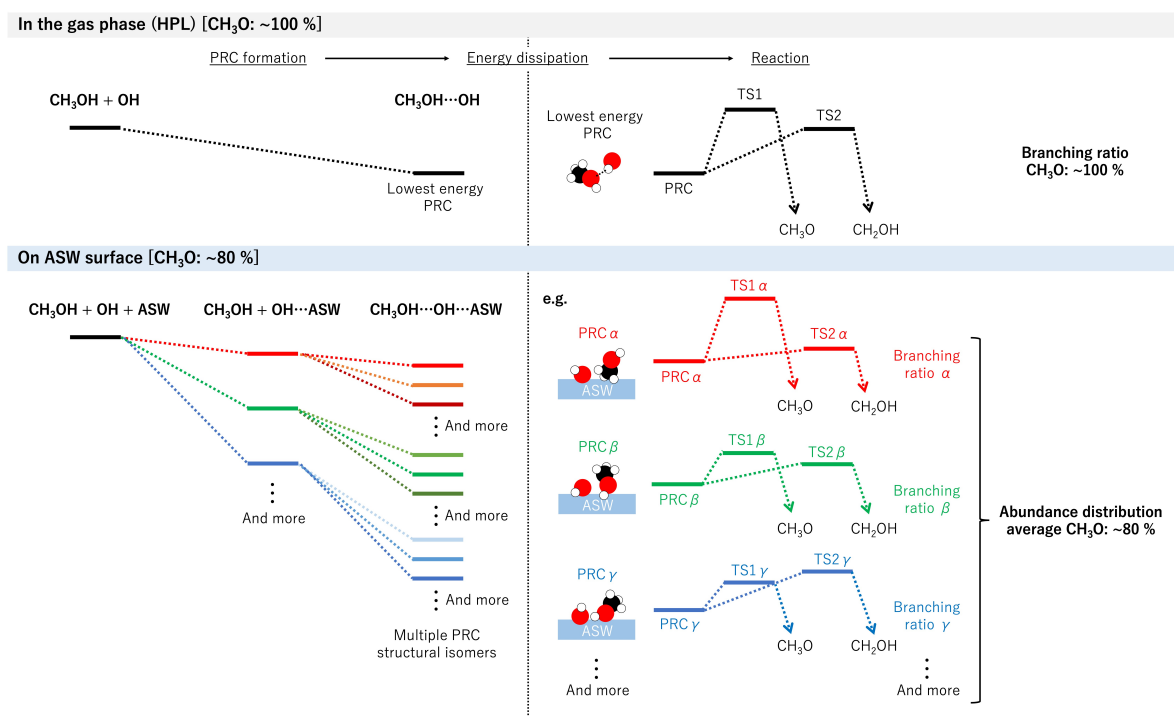

Figure 7. Schematic diagram of potential energy surfaces for the $CH_3OH + OH$ reaction with reference to previous calculations [Xu & Lin 2007, Shannon et al. 2013, Gao et al. 2018, Jara-Toro et al. 2017, Sameera et al. 2023]. In the gas phase (HPL), only the lowest energy PRC is considered, and the branching ratio of $CH_3O$ is ~100% at low temperature (T < 80 K) [Gao et al. 2018]. On the ASW surface, because multiple PRC structural isomers involving many $H_2O$ molecules (PRC α, β, γ, and others) can exist stably at low temperature, they have different potential barrier shapes. Therefore, the obtained branching ratio of ~80% $CH_3O$ in the present experiment represents the average abundance distribution of the branching ratios obtained from each PRC. Note that each multiple PRC structural isomer has a different absolute energy value. The descriptions of $CH_3OH…OH$, $OH…ASW$, and $CH_3OH…OH…ASW$ indicate formation of a complex (or adsorption).

Although it is generally known that tunneling reactions often bring large isotopic effects, little isotopic effects were detected on the branching ratio on the ASW surface. In the HPL gas-phase calculations at T <70 K, because the formation of the PRC preceding the hydrogen abstraction from methanol by OH was the rate-determining process [Gao et al. 2018], the isotopic effect on the overall abstraction reaction rate constant (i.e., the sum of methoxy and hydroxyethyl formation reactions) was almost negligible. In the present experiment, the sum of the rate constants for the formation of methoxy and hydroxymethyl radicals (i.e., $k_{1+2}$) shows no substantial isotopic effect, which is presumably because the association process between methanol and OH (i.e., adsorption, diffusion, and PRC formation process) is the rate-determining process of the reaction. In fact, it has been reported that when tunneling reaction rate on the cold surface is limited by surface diffusion, a significant isotope effect does not appear [Hama et al 2015].

However, our result of no substantial isotopic effects for branching ratio (i.e., $k_1/k_2$) on ASW could not be explained solely by the above reasons. In the gas-phase, because the branching ratio for the formation of $CH_3O$ or $CH_2OH$ was determined by the competition between the two tunneling reactions after PRC formation, isotopic effects could affect the branching ratio in the $CD_3OH$ and $CH_3OD$ cases. This is because different isotopic atom (i.e., H or D) abstraction reactions occur from the methyl and hydroxyl groups, respectively, which could change their competing reaction rate ratios. According to the HPL calculation, the rate constant of $CH_3O$ formation has been reported to be 3–4 orders of magnitude larger than that of $CH_2OH$ formation at low temperatures, where the branching to $CH_3O$ is almost unity [Gao et al. 2018]. In such cases, even if there are large isotopic effects on the reactions, the formation of methoxy radical could remain overwhelmingly dominant in each isotopologue experiment, and the isotopic effect may not appear in the branching ratio detected experimentally. However, our results would conflict with the above consideration, which is suggested to explain the reason for the lack of isotopic effects because the substantial formation of hydroxymethyl radicals (same order of methoxy formation, methoxy:hydroxymethyl ~4:1) was observed. It is difficult to understand the observed branching ratio on ASW from the analogy in the gas-phase reactions using a single PRC. The speculation that the branching ratio is determined by the sum of the contributions of multiple PRC structures could explain the lack of isotopic effects in the branching ratios measured on ASW. In other words, if methoxy radical production is overwhelmingly dominant in some types of PRCs such as PRC $\gamma$ in Fig. 7 and hydroxymethyl radical production is overwhelmingly dominant in other types of PRCs such as PRC $\alpha$ in Fig. 7, and the ratio of the presence of these two types of PRCs broadly classified is ~4:1, the branching ratio (methoxy/hydroxymethyl) of ~4 could be obtained without experimentally detectable isotopic effects.

## 3.4. Diffusion of Methoxy Radical

Fig. 4 clearly shows that formaldehyde was formed by consuming two OH radicals. It means that the methoxy and/or hydroxymethyl radicals diffuse significantly to encounter secondary OH, because the coverage of OH radicals on UV-irradiated ASW is very small (~ 1%). Therefore, it is expected that the contributions of formaldehyde formation at each reaction pathway (see Fig. 6(a)) strongly depend on the diffusivity of each methoxy and hydroxymethyl radical. Fig. 8 shows a comparison of the time variation of pickup intensities for methoxy and hydroxymethyl radicals obtained (a) $CD_3OH$ and (b) $CH_3OD$ experiments, respectively. To facilitate the comparison of formaldehyde formation contributions through methoxy and hydroxymethyl radicals, the intensities of hydroxymethyl radicals in that figure are adjusted by multiplying them with the branching ratio ($k_1/k_2$ ~ 4.3). The discrepancy between the time variations of $k_1/k_2 \times [CH_2OH]_t$ and $[CH_3O]_t$ should appear significantly at the long deposition time because it is attributed to the difference of consumption terms of $CH_2OH$ and $CH_3O$ (see Eq. 12 and 13), respectively. In Fig. 8, the discrepancy in the signal intensity between methoxy and hydroxymethyl radicals at each isotopologues experiments significantly appears at above 500 s, and the intensities of methoxy radicals tend to be smaller than that of hydroxymethyl radicals with time. This implies that the contributions of consumption by the sequential reaction with the diffusion of methoxy radicals are more significant than that of hydroxymethyl radicals in both isotopologues experiments. The adsorption energies of $CH_3O$ and $CH_2OH$ on ASW were calculated to be 0.32 eV [Sameera et al. 2021] and 0.46 eV [Sameera et al. 2023] as the average value for several adsorption sites. The value of $CH_2OH$ is significant, considering the value of $CH_3OH$ is 0.39 eV [Sameera et al. 2023]. Thus, the relation of diffusivity can be concluded to $CH_3O > CH_3OH > CH_2OH$ under the simple assumption that adsorption energy positively correlates to the activation energy of diffusion. This finding supports the results in Fig. 6 that the pickup intensity of HDO can be represented using a linear function, meaning neglect of reaction (11), in $CH_3OD$ experiments, although a quadratic function was required to fit that of HDO in $CD_3OH$ experiments.

Additionally, we can discuss the difference in the signal intensity between $CH_3O$ and $CD_3O$ after 500 s in Fig. 2c,d. This difference may be due to isotopic effects appeared in abstraction reaction and/or diffusion of methoxy radicals in the methoxy consumption reactions (7) and (8). The consumption of $CH_3O$ (i.e., reaction (7)) should be larger than that of $CD_3O$ (i.e., reaction (8)), because the abstraction of D atoms tends to be slower than that of H atoms, and $CD_3O$ should also tend to be slower than $CH_3O$ due to the non-identical binding energy on ASW by the zero-point vibrational energy difference ($k_7 > k_8$). In the case of hydroxymethyl radicals, the fact that little isotope effect appears after ~500 s may be due to the small contributions of their sequential reactions (10) and (11) ($k_{10} \sim k_{11} \sim 0$).

Unfortunately, the present experiments could not reveal the mechanism of methoxy radical diffusion. It is speculated that the radical could diffuse only when trapped in particularly weak adsorption sites following reaction formation, or diffuse transiently using the heat of reaction.

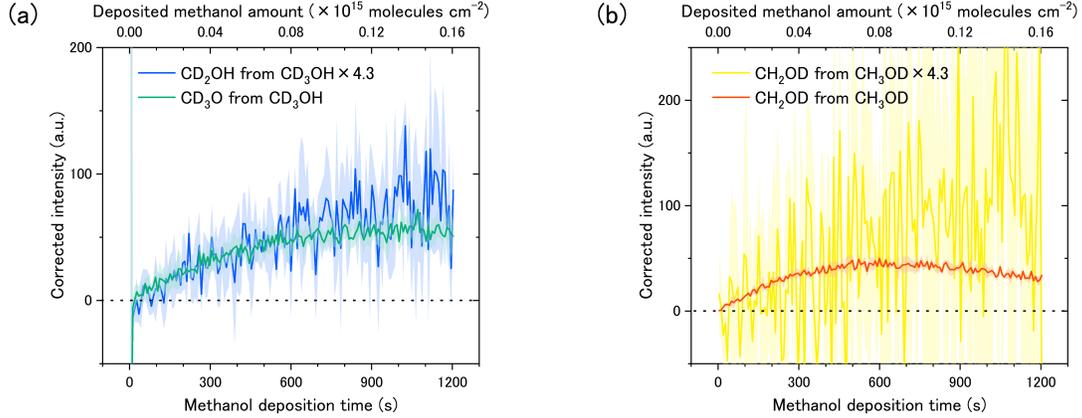

Figure 8. Comparisons of the trend of time variation of methoxy and hydroxymethyl (multiplied by $k_1/k_2 = 4.3$) radicals for (a) $CD_3OH$ and (b) $CH_3OD$ experiments to examine the degree of the contribution for the consumption reactions.

4. Astrophysical Implications and Conclusions

Astronomical observations showed that the abundances of $CH_2OH$-bearing COMs (i.e., $HCOCH_2OH$ and $(CH_2OH)_2$) in the gas phase tend to be smaller than those of $CH_3O$-bearing COMs (i.e., $HCOOCH_3$, $CH_3OCH_3$, and $CH_3OCH_2OH$) [Jørgensen et al. 2020; Mininni et al. 2020; Manigand et al. 2020; El-Abd at al. 2019; Tercero et al. 2018; Rivilla et al. 2017; Coutens et al. 2015; Taquet et al. 2015]. Although gas-phase reactions have also been proposed for these formation pathways [e.g., Balucani et al. 2015], the roles of radicals on dust still need to be evaluated. The exposure of CO and $CH_3OH$ solid mixtures to UV and/or H atoms have shown that the yields of $CH_3O$-bearing COMs are inefficient compared to gas-phase observations [Chuang et al. 2017]. However, the formation of $CH_3O$-bearing COMs was found to be enhanced when $CH_3OH$ coexists with water ice [Ishibashi et al. 2021]. It was proposed that the reaction of methanol with OH radicals produced by photodissociation of water efficiently produce $CH_3O$ on ice. The new dust model of Kouchi et al. 2021a,b indicates that CO cannot fully cover water ice mantle and thus coexistence of $CH_3OH$ and $H_2O$ at the sites of photolysis would be very likely. The present experiment quantitatively shows that the reaction of $CH_3OH$ with OH on ASW clearly can make $CH_3O$-rich condition. That is, the reaction would increase the concentration of $CH_3O$ and lead to enhancement in the abundance of $CH_3O$-bearing COMs in the ice, which in turn sublimate and enrich the gas phase.

The present reactions should also take place at high temperatures, where methanol and OH can thermally diffuse [Furuya et al. 2022; Miyazaki et al. 2022]. In these high-temperature regions, because methanol would tend to form the most stable adsorption structure [Dawes et al. 2016], the formation of $CH_3O$ through the H atom abstraction from the hydroxyl group could be enhanced.

The present results may provide information for updating chemical models of COMs formation. The present value of the branching ratio ($CH_3O/CH_2OH$ = 4.3 ± 0.6) can also be applied to all methanol isotopologues, as the isotopic effects were little observed in our experiments. The effective reaction rate constants on the dust could depend on their diffusion/association mechanism, since this reaction may be rate-determining in the process of association between methanol and OH (i.e., PRC formation). The present results also suggest that branching due to chemical reactions on the dust should be statistically considered not only for specific adsorption structures but also for various adsorption structures.

In addition, the present experiments suggest that even at 10 K, methoxy radicals produced by the reaction of $CH_3OH$ and OH can diffuse to some extent on the ice surface because the sequential reactions to produce formaldehyde require the diffusion of methoxy radicals to encounter another OH under OH-poor conditions. Therefore, when $CH_3O$ is produced on interstellar dust, it may cause diffusive reactions with other molecules and radicals even in cold environments, which could lead to efficient formation of $CH_3O$-derived COMs. Unfortunately, details of the diffusion mechanism cannot be clarified in the present experiments. Because $CH_3O$ diffusion even at 10 K crucial in chemical models on dust in cold molecular clouds, the diffusion mechanism, including diffusion distance, needs to be further investigated in the future.

## 5. Acknowledgments

This work was supported by JSPS Grant-in-Aid for Scientific Research (JP22H00159). The authors thank the staff of Technical Division at Institute of Low Temperature Science for making various experimental devices.

## Appendix

### A1. No Effect of $Cs^+$ Ion Irradiation on the Surface Reaction in the Present Experimental System

We investigated the effect of $Cs^+$ ion irradiation on the formation of $CH_3O$, $CH_2OH$, and $H_2CO$. Fig. 9 shows the pickup intensities of the products obtained in the $CH_3OH$ experiments, in which $Cs^+$ irradiation was delayed by ~300 s. The pickup signal intensities were consistent with those in the experiment without the delay of $Cs^+$ irradiation (solid lines), indicating that the effect of $Cs^+$ ion irradiation on the surface reaction was negligible.

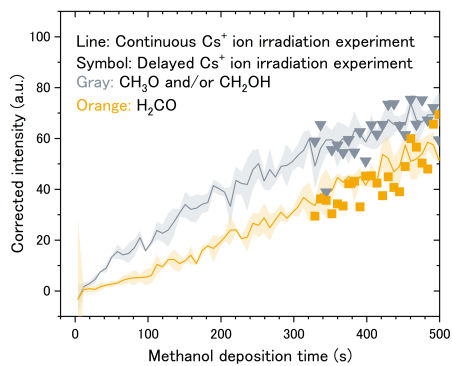

Figure 9. Variations in the time evolution of $CH_3O$ and/or $CH_2OH$ (gray) and $H_2CO$ (orange) signals during $CH_3OH$ deposition using different $Cs^+$ ion irradiation procedures: (solid lines) continuous $Cs^+$ irradiation and (symbols) experiments with $Cs^+$ irradiation delayed for ~300 s.

## A2. Contributions from Contaminants with the Same Mass as the Reaction Product

The pickup signal of the target species (e.g., $CD_3O$ in the $CD_3OH$ experiments) would have contributions from that of other species at the same mass (Table 1), as mentioned in the main text. For example, the pickup signal at mass 34 in the $CD_3OH$ experiments would have multiple contributions from the $CHD_2OH$ originally present in the chemical reagent of $CD_3OH$, $H_2O_2$ generated by the photolysis of ASW, and $CD_3O$ formed by $CD_3OH$ + OH. In addition, the pickup signal at mass 33 would have contributions from $HO_2$ generated from UV-irradiated ASW and $CD_2OH$ formed by $CD_3OH$ + OH. The pickup signal at mass 32 in the $CH_3OD$ experiments would have contributions from $CH_3OH$ generated by the D–H substitution reaction in the gas line and/or originally present in the $CH_3OD$ reagent. For HDO (mass = 19), $H_2^{17}O$ and HDO contamination from the gas line ($HDO_{GL}$) would be possible candidates for interference. In the present study, the contaminant amounts (except for $H_2O_2$ and $HO_2$) were estimated from blank experiments with no UV irradiation. The amounts of $H_2O_2$ and $HO_2$ were estimated from the $CH_3OH$ experiments.

Table 1. Masses of target species and the contaminants in each experiment.

| Experiments | Mass | Target species | Contaminants |
|---|---|---|---|
| **$CD_3OH\_uv$** | 19 | HDO | $H_2^{17}O$, $HDO_{GL}$ |
| | 33 | $CD_2OH$ | $HO_2$ |
| | 34 | $CD_3O$ | $CHD_2OH$, $H_2O_2$ |
| **$CD_3OH\_blank$** | 19 | | $H_2^{17}O$, $HDO_{GL}$ |
| | 34 | | $CHD_2OH$ |
| **$CH_3OH\_uv$** | 33 | $HO_2$ | $^{13}CH_3OH$ |
| | 34 | $H_2O_2$ | $CH_3^{18}OH$ |
| **$CH_3OH\_blank$** | 33 | | $^{13}CH_3OH$ |
| | 34 | | $CH_3^{18}OH$ |
| **$CH_3OD\_uv$** | 19 | HDO | $H_2^{17}O$, $HDO_{GL}$ |
| | 32 | $CH_2OD$ | $CH_3OH$ |
| **$CH_3OD\_blank$** | 19 | | $H_2^{17}O$, $HDO_{GL}$ |
| | 32 | | $CH_3OH$ |

## A3. Methods for Correcting the Pickup Signal of Products

In the Cs⁺ ion pickup method, it is difficult to maintain collection efficiency unity between experiments because of technical issues. Hence, to remove the influence on the fluctuation of the collection efficiency, the obtained raw data needed to be corrected for further evaluation. Figs. 10(a) and (b) show the time variations in the raw pickup signal intensities ($I_x(t)$) at masses 34 and 35, respectively, after UV irradiation for three different $CD_3OH$ experiments. The signal intensity at mass 34 was composed of contributions from three species: $CD_3O$, $H_2O_2$, and $CHD_2OH$. The signal intensity appearing before methanol deposition is the contribution of $H_2O_2$ formed by pre-UV irradiation to ASW. To correct the variations in $I_x(t)$ between each experiment, $I_x(t)$ was corrected using the following equation:

$$I_{corrected\_x}(t) = I_x(t)/I_{methanol}(t) \times I_{Standard\_methanol}(t)$$

where $I_{methanol}(t)$ is the raw signal intensity of methanol and $I_{Standard\_methanol}(t)$ is the standard signal intensity of methanol. $I_{Standard\_methanol}(t)$ can be expressed by the raw $H_2O$ signal intensities before methanol deposition ($I_{H2O}$) and the raw methanol signal intensities before UV irradiation ($I_{blank\_methanol}(t)$) as follows:

$$I_{Standard\_methanol}(t) = I_{blank\_methanol}(t)/I_{H2O} \times 10^4$$

Note that $I_{Standard\_methanol}(t)$ is unique for each methanol isotopologue because of differences in the pickup efficiency. This data correction procedure indicates that the signal intensities of all species are constructed using a value of $10^4$ cps, which is defined as the standard signal intensity of $H_2O$ on ASW before UV irradiation. Variations in the value of $I_{corrected\_x}(t)$ with time were consistent in the three experiments (Fig. 10(c)), indicating that the relative intensity did not vary between every experiment. The signal variations between each experiment were mainly ascribed to the variation in the collection efficiency of the analyzer and were not problematic for achieving quantitative chemical compositions on ASW. Hence, $I_{corrected\_x}(t)$ would be suitable for the quantitative analysis in the present experiment.

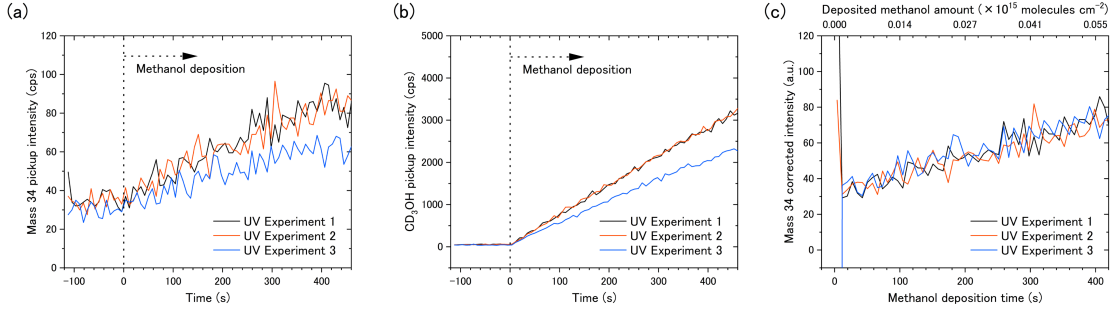

Figure 10. The correction procedure (e.g., mass 34 signal in CD$_3$OH experiments): (a) Mass 34 pickup signals: $I_x(t)$, (b) CD$_3$OH (mass 35) pickup signals: $I_{methanol}(t)$, (c) corrected pickup signal of mass 34: $I_{corrected\_x}(t)$.

A4. Derivation of Rate Equations for Reactions Occurring in the Present Experiment and the Time-dependent Equation of the Product Concentrations

First, we derived rate equations for reactions (1, 2, 7, and 9). For reactions (1) and (2) to occur, the vapor-deposited methanol must associate with OH on the ASW surface. The following two cases are considered for the association process between methanol and OH: (i) methanol associates with OH during transient diffusion using its adsorption energy, and (ii) thermalized methanol and OH adsorbed on ASW associate via their thermal diffusion. On the ASW at 10 K, because methanol and OH would not diffuse well on a laboratory time scale [Furuya et al. 2022; Miyazaki et al. 2022], most association reactions should occur via the process (i). Under the assumption that methanol, which does not contact OH before the thermalization, does not subsequently contribute to any chemical reactions, variations in the surface number density of OH via reactions (1) and (2) can be approximately expressed as a first-order reaction, such as the photodissociation reaction systems, and are described by the following rate equations:

$$\frac{d[OH]_t}{dt} = -(k_1 + k_2)[OH]_t - k_7[CH_3O]_t[OH]_t - k_9[CH_2OH]_t[OH]_t, \quad (A1)$$

$$\frac{d[CH_3O]_t}{dt} = k_1[OH]_t - k_7[CH_3O]_t[OH]_t, \quad (A2)$$

$$\frac{d[CH_2OH]_t}{dt} = k_2[OH]_t - k_9[CH_2OH]_t[OH]_t, \quad (A3)$$

$$\frac{d[H_2CO]_t}{dt} = k_7[CH_3O]_t[OH]_t + k_9[CH_2OH]_t[OH]_t, \quad (A4)$$

where $k_1$ (also $k_2$) can be described as follows:

$$k_1 \ [s^{-1}] = \sigma_1 \ [cm^2 \ molecule^{-1}] \times f_{methanol} \ [molecule \ cm^{-2} \ s^{-1}],$$

where $\sigma_1$ is the reaction cross-section (including the association process) of methanol and OH (reaction 1) and $f_{\text{methanol}}$ is the flux of methanol deposition.

Next, from Eqs. (A1)–(A4), the variations in the concentrations of the reaction products were obtained. For simplicity, the second and third terms in Eq. (A1) were ignored because of their negligible contributions to the reaction rate, providing the following equations:

$$[OH]_t = [OH]_0 \, e^{-k_{1+2}t}, \quad (A5)$$

$$[CH_3O]_t = \frac{k_1}{k_7}\left[1 - exp\left\{-\frac{k_7}{k_{1+2}}[OH]_0(1 - e^{-k_{1+2}t})\right\}\right], \quad (A6)$$

$$[CH_2OH]_t = \frac{k_2}{k_9}\left[1 - exp\left\{-\frac{k_9}{k_{1+2}}[OH]_0(1 - e^{-k_{1+2}t})\right\}\right], \quad (A7)$$

$$[H_2CO]_t = [OH]_0(1 - e^{-k_{1+2}t}) - \frac{k_1}{k_7}\left[1 - exp\left\{-\frac{k_7}{k_{1+2}}[OH]_0(1 - e^{-k_{1+2}t})\right\}\right] - \frac{k_2}{k_9}\left[1 - exp\left\{-\frac{k_9}{k_{1+2}}[OH]_0(1 - e^{-k_{1+2}t})\right\}\right]. \quad (A8)$$

Furthermore, the Taylor expansion $\left(exp\left\{-\frac{k_7 \text{ or } k_9}{k_{1+2}}[OH]_0(1 - e^{-k_{1+2}t})\right\}\right)$ in Eqs. (A6)–(A8) up to the second-order term provides approximations for Eqs. (12)–(14) in the main text. Based on the experimental results (see main text), it is reasonable to assume that the second and third terms in Eq. (A1) can be ignored.

References

Antiñolo, M., Agúndez, M., Jiménez, E., et al. 2016, *Astrophys. J.* **823**, 25

Bacmann, A., Taquet, V., Faure, A., et al. 2012, *Astron. Astrophys.* **541**, L12

Balucani N., Ceccarelli C., and Taquet V. 2015, *Mon. Not. R. Astron. Soc.* **449**, L16

Canosa, A. 2019, *Proceedings of the International Astronomical Union,* *15*(S350), 35-40

Cernicharo, J., Marcelino, N., Roueff, E., et al. 2012, *Astrophys. J. Lett.* **759**, L43

Chang, Q., & Herbst, E. 2016, *Astrophys. J.* **819**, 145

Chao, W., Lin, J. J., Takahashi, K., et al. 2019, *Angew. Chem. Int. Ed.* **58**, 5013; 2019, *Angew. Chem.* **131**, 5067

Chuang, K. J., Fedoseev, G., Ioppolo, S., et al. 2016, *Mon. Not. R. Astron. Soc.* **455(2)**, 1702-1712.

Chuang, K. J., Fedoseev, G., & Qasim, D. 2017, *Mon. Not. R. Astron. Soc.* **467**, 2552

Coutens, A., Persson, M. V., Jørgensen, J. K., et al. 2015, *Astron. Astrophys.* **576**, A5

Dawes, A., Mason, N. J., & Fraser, H. J. 2016, *Phys. Chem. Chem. Phys.* **18**, 1245

del Mazo-Sevillano, P., Aguado, A., Jiménez, E., et al. 2019, *J. Phys. Chem. Lett.* **10**, 1900

El-Abd, S. J., Brogan, C. L., Hunter, T. R., et al. 2019, *Astrophys. J.* **883**, 129

Fedoseev, G., Cuppen, H. M., Ioppolo, S., et al. 2015, *Mon. Not. R. Astron. Soc.* **448(2)**, 1288-1297

Ferrero, S. Zamirri, L., Ceccarelli, C., et al. 2020, *Astrophys. J.* **904**, 11

Furuya, K., Hama, T., Oba, Y., et al. 2022, *Astrophys. J. Lett.* **933**, L16

Gao, L. G., Zheng, J. J., Fernández-Ramos, A., et al. 2018, *J. Am. Chem. Soc.* **140**, 2906

Garrod, R. T. & Herbst, E. 2006, *Astron. Astrophys.* **457**, 927

Gómez Martín, J. C., Caravan, R. L., Blitz, M. A., et al. 2014, *J. Phys. Chem. A* **118**, 2693

Hama, T., Ueta, H., Kouchi, A., & Watanabe, N. 2015, *Proceedings of the National Academy of Sciences, 112(24), 7438-7443*

He, J., Simons, M., Fedoseev, G., et al. 2022 *Astron. Astrophy.* **659**, A65

Hidaka, H., Watanabe, M., Kouchi, A., & Watanabe, N. 2009, *Astrophys. J.* **702**, 291

Ishibashi, A., Hidaka, H., Oba, Y., at al. 2021, *Astrophys. J. Lett.* **921**, L13

Jara-Toro, R. A., Hernández, F. J., Taccone, R. A., et al. 2017, *Angew. Chem. Int. Ed.* **56**, 2166; 2017, *Angew. Chem.* **129**, 2198

Jiménez-Serra, I., Vasyunin, A. I., Caselli, P., et al. 2016, *Astrophys. J. Lett.* **830**, L6

Jin, M., & Garrod, R. T. 2020, *Astrophys. J.* **249**, 26

Jørgensen, J. K., Belloche, A., & Garrod, R. T. 2020, *Annu. Rev. Astron. Astrophys.* **58**, 727

Kang, H. 2011, *Bull. Korean Chem. Soc.* **32**, 389

Kouchi, A., Tsuge, M., Hama, T., et al. 2021a, *Mon. Not. R. Astron. Soc.* **505**, 1530

Kouchi, A., Tsuge, M., Hama, T., et al. 2021b, *Astrophys. J.* **918**, 45

Manigand, S., Jørgensen, J. K., Calcutt, H., et al. 2020, *Astron. Astrophys.* **635**, A48

Mininni, C., Beltrán, M. T., Rivilla, V. M., et al. 2020, *Astron. Astrophys.* **644**, A84

Miyazaki, A., Tsuge, M., Hidaka, H., et al. 2022, *Astrophys. J. Lett*. **940**, L2

Miyazaki, A., Watanabe, N., Sameera, W. M. C., et al. 2020, *Phys. Rev. A* **102**, 052822

Nagaoka, A., Watanabe, N., & Kouchi, A. 2005, *Astrophys. J. Lett.* **624**, L29

Nagaoka, A., Watanabe, N., & Kouchi, A. 2007, *J. Phys. Chem. A* **111**, 3016

Nguyen, T. L., Ruscic, B., & Stanton, J. F. 2019, *J. Chem. Phys.* **150**, 084105

Öberg, K. I., Bottinelli, S., Jørgensen, J. K., & van Dishoeck, E. F. A. 2010, *Astrophys. J.* **716**, 825

Öberg, K. I., Garrod, R. T., van Dishoeck, E. F., & Linnartz, H. 2009, *Astron. Astrophys.* **504**, 891

Ocaña, A. J., Blázquez, S., Potapov, A., et al. 2019, *Phys. Chem. Chem. Phys.* **21**, 6942

Paardekooper, D. M., Bossa, J. B., & Linnartz, H. 2016, *Astron. Astrophys.* **592**, A67

Rivilla, V. M., Beltrán, M. T., Cesaroni, R., et al. 2017, *Astron. Astrophys.* **598**, A59

Roncero, O., Zanchet, A., & Aguado, A. 2018, *Phys. Chem. Chem. Phys.* **20**, 25951

Sameera, W. M. C., Jayaweera, A. P., Ishibashi, A. et al. 2023, *Faraday Discuss.* Accepted Manuscript DOI: 10.1039/D3FD00033H

Sameera, W. M. C., Senevirathne, B., Andersson, S. et al. 2021, *J. Phys. Chem. A* **125**, 1, 387-393

Shannon, R. J., Blitz, M. A., Goddard, A., & Heard, D. E. 2013, *Nat. Chem.* **5**, 745

Shingledecker, C. N., Tennis, J., Gal, R. L., & Herbst, E. 2018, *Astrophys. J.* **861**, 20

Slanger, T. G., & Black, G. 1982, *J. Chem. Phys.* **77**, 2432-2437

Soma, T., Sakai, N., Watanabe, Y., & Yamamoto, S. 2018, *Astrophys. J.* **854**, 116

Song, L., & Kästner, J. 2017, *Astrophys. J.* **850**, 118

Taquet, V., Ceccarelli, C., & Kahane, C. 2012, *Astron. Astrophys.* **538**, A42

Taquet, V., López-Sepulcre, A., Ceccarelli, C., et al. 2015, *Astrophys. J.* **804**, 81

Tenelanda-Osorio, L. I., Bouquet, A., Javelle, T., et al. 2022, *Mon. Not. R. Astron. Soc.* **515**, 5009

Tercero, B., Cuadrado, S., López, A., et al. 2018, *Astron. Astrophys.* **620**, L6

Vastel, C., Ceccarelli, C., Lefloch, B., & Bachiller, R. 2014, *Astrophys. J. Lett.* **795**, L2

Watanabe, N., & Kouchi, A. 2002, *Astrophys. J. Lett.* **571**, L173

Wu, J., Gao, L. G., Varga, Z., et al. 2020, *Angew. Chem. Int. Ed.* **59**, 10826; 2020, *Angew. Chem.* **132**, 10918

Xu, S., & Lin, M. C. 2007, *Proc. Combust. Inst.* **31**, 159

Yocum K., Milam S., Gerakines P., & Weaver S. W. 2021, *Astrophys. J.* **913**, 61